\documentstyle[12pt,preprint]{aastex}

\begin{document}

\title{A High-resolution Spectrum of the R CrB Star V2552 Ophiuchi} 
\author{
N. Kameswara Rao}
\affil{Indian Institute of Astrophysics;
Bangalore,  560034 India\\
nkrao@iiap.ernet.in}


\author{David L. Lambert}
\affil{Department of Astronomy; University of
Texas; Austin, TX 78712-1083
\\ dll@astro.as.utexas.edu}

\begin{abstract}
Photometry and low-resolution spectroscopy have
added V2552 Oph to the rare class of R Coronae
Borealis variables. We confirm this classification
of V2552 Oph through a comparison of our high-resolution
optical spectrum of this star and that of R CrB and other F-type
members
of the class. We show that V2552 Oph most closely resembles
Y Mus and FH Sct, stars in which Sr, Y, and Zr are enhanced.

{\it Subject headings: stars:variables:other (R CrB)}

\end{abstract}

\section{Introduction}

The R Coronae Borealis variables are a select group of distinctive
supergiants. A Galactic census presently recognizes no more than
about  30  R CrB stars. Their especial
distinctions are a severe deficiency of hydrogen in their
atmosphere, and a propensity to fade by up to 6 - 8
magnitudes at irregular and unpredictable times. 
The origin of the hydrogen deficiency (and other composition
anomalies) and the mechanism responsible for the characteristic
fading of the star remain the subjects of theoretical and
observational enquiries. Given the small number of R CrB
stars, one seeks to enlarge the sample in the hope that
a new discovery may yield a fruitful clue to outstanding
questions.

The latest addition to the R CrB census  is V2552 Oph (Kazarovets et al.
2003). This star CD -22$^\circ$\,12017 was identified as a R CrB
star from a light curve showing two R CrB-like fadings (Kato \&
Katsumi 2003).  This identification was confirmed by
Hesselbach, Clayton, \& Smith (2002) from photometry and a low
resolution spectrum. In this paper, we report on the first high-resolution
optical spectrum of V2552 Oph, and show that it is remarkably
similar to the spectrum of the eponym with a few striking differences. 

\section{Observations}

Our spectrum of V2552 Oph was obtained on June 11, 2003 (JD 2452801.809)
 with the  McDonald
Observatory's
 2.7m telescope and the
{\it 2dcoud\'{e}} echelle spectrograph (Tull et al. 1995\markcite{Tul95})
at a resolving power of 60,000. The spectrum runs from about 3800 \AA\
to 10,000 \AA\ with gaps between the orders beyond 5800 \AA.
The signal-to-noise (S/N) ratio in the continuum is 95 or greater
over much of the observed wavelength interval longward of 5500\AA. 
The S/N ratio decreases to shorter wavelengths.
V2552 Oph
appeared to be at maximum light. 
 Standard reductions of the CCD frames
give the radial velocity of V2552 Oph as 60.5$\pm0.9$ km s$^{-1}$
from a set of 51 unblended photospheric lines.

\section{V2552 Oph and R CrB: Spectra and Abundances}

By inspection of the spectra, it was apparent that V2552 Oph has a spectrum
remarkably similar to that of R CrB. This is illustrated in
Figures 1  and 2.  The similarity suggests that the stars
have similar atmospheres and compositions.  
Lines which are of very similar strength in V2552 Oph and R CrB
belong to the following atoms and ions: H\,{\sc i}, C\,{\sc i},
Na\,{\sc i}, Mg\,{\sc i}, Al\,{\sc i}, Si\,{\sc i}, Si\,{\sc ii},
S\,{\sc i}, K\,{\sc i}, Ca\,{\sc i}, Sc\,{\sc ii}, Ti\,{\sc ii},
Fe\,{\sc i}, Fe\,{\sc ii}, Ni\,{\sc i}, Zn\,{\sc i}, Ba\,{\sc ii}, and
La\,{\sc ii}. The similarity for C\,{\sc i} is a direct consequence
of the fact that the neutral carbon atom is the principal contributor
of continuous opacity (Sch\"{o}nberner 1975).
 The similarity for other atoms and ions with a
mix of low to high excitation lines is an indication that the
effective temperature ($T_{\rm eff}$), surface gravity ($g$),
microturbulence ($\xi_t$), and composition are also
similar for the pair of stars.

Obvious differences in line strength for atoms and ions are noted  
which imply differences in abundance: 
\begin{itemize}
\item
 The Li\,{\sc i} 6707 \AA\ resonance doublet is strong in
R CrB but absent from V 2552 Oph (Figure 1).
\item
The N\,{\sc i} lines, all of high-excitation, are considerably
strengthened in V2552 Oph. Figure 1 shows the N\,{\sc I} lines
at 6706.1\AA\ and 6708.7\AA\ to be enhanced in V2552 Oph relative
to R CrB.
\item
The O\,{\sc i} high excitation and the low excitation
6363\AA\ [O\,{\sc i}] lines are
weaker in V2552 Oph. The fact that the low and high excitation are
all weaker suggests that the oxygen abundance is lower for V2552 Oph. 
\item
Lines of Sr\,{\sc ii}, Y\,{\sc ii}, and Zr\,{\sc ii} are
much stronger in V2552 Oph (Figure 2). 
\end{itemize}
The lines of the C$_2$ Swan bands are much weaker in V2552 Oph than in
R CrB. The rotational excitation of the C$_2$ molecules is
representative of the photosphere.

An abundance analysis of V2552 Oph and R CrB was made using the family of
model atmospheres employed by Asplund et al. (2000). The R CrB spectrum
is one obtained in 1995 August with the {\it 2dcoud\'{e}} 
at the same spectral resolution as our spectrum of V2552 Oph. In light of
the spectral similarities, we adopted as the initial
(and final) model one
with the atmospheric parameters chosen by Asplund et al. for
R CrB: $T_{\rm eff} = 6750$ K, $\log g$ = 0.5 cgs, and
$\xi_t = 7$ km s$^{-1}$, and an assumed ratio C/He = 0.01
by number of atoms.
Standard spectroscopic checks for $T_{\rm eff}$, $g$, and $\xi_t$
confirm the adopted parameters to be appropriate for V2552 Oph.
 Results of the
abundance analysis are given in Table 1 where we list results
for V2552 Oph and R CrB from our spectra, and results for
R CrB from Asplund et al. Selection of lines and atomic
data (i.e., the $gf$-value of a line) are taken largely from
Asplund et al. For a few Fe\,{\sc i} lines, $gf$-values were taken
from Reddy et al. (2003).
 In general,
the same lines were used for both stars and, hence,  the
abundance analyses are  accurate in a differential
sense. There is very good agreement between the two R CrB analyses;
the difference  seen for oxygen and barium
is due to an improved selection of lines in the case of
oxygen and to an erroneous equivalent width for one Ba\,{\sc ii}
line used in  Asplund et al.'s analysis.

Table 1 shows that for many elements there is an
uncanny match of the abundances for V2552 Oph and R CrB:
differences are 0.05 dex or less in many cases. 
This degree of similarity is surely fortuitous.
Abundance differences for N, Y, and Zr average
about 0.6 dex in favor of V2552 Oph and the difference for O
is 0.6 dex in favor of R CrB.

\section{V2552 Oph among the R CrBs}

If classified by composition, the F-G type  R CrB stars belong to
one of two classes - a majority and a minority class (Lambert
\& Rao 1994). The minority R CrB stars are marked by an
approximately solar Si abundance and a severe Fe deficiency or,
equivalently, an
 extraordinarily high Si/Fe
 - see Lambert \& Rao (1994)
and Asplund et al. (2000) for a fuller characterization of the
minority class. Majority members have a  solar or
slightly subsolar Si abundance
and a mild  Fe deficiency resulting in a less unusual Si/Fe ratio ([Si/Fe]
 $\simeq$
+0.5).  V2552 Oph with [Si/Fe] = $+$0.5 and a 1 dex Fe underabundance
becomes the newest member of the majority. 

Within the majority, there are significant abundance differences for
some elements, notably those few elements for which V2552 Oph and R CrB
differ: Li, N, O, Sr, Y, and Zr. Except for a coupling of Sr, Y, and
Zr, these elements do not vary in a correlated manner.
Among majority R CrBs previously analysed, Y Mus and FH Sct closely
resemble V2552 Oph in composition across the elements from H to La.
In particular, the trio have no detectable Li, similar low abundances of O,
and high Sr, Y, and Zr abundances (relative to Fe and to Ba and La). 
We refer the reader to Asplund et al. (2000) for a discussion 
of what may be learnt from the compositions about the origins
of these variable stars.

\section{Interstellar and circumstellar gas}

Absorption components of likely interstellar origin are present in the
Na D lines (Figure 3). Stellar Na D lines at +60 km s$^{-1}$ are accompanied
by strong  components extending from -10.5 to 0.0
km s$^{-1}$. The strongest components are seen in the K\,{\sc i}
7699\AA\ line at -7.1 and -0.6 km s$^{-1}$.
Diffuse interstellar bands are present in V2552 Oph's spectrum
at 5780, 6284, 6379, and 6613\AA. Adopting the wavelengths
given by Herbig (1995), we calculate a heliocentric velocity of
about -7 km s$^{-1}$ from these bands. 
Herbig's (1993) relation between the equivalent width of a
diffuse interstellar band and reddening gives E$_{\rm B-V}$ = 0.4
for V2552 Oph.

This estimate, the apparent visual magnitude (V = 10.7, Kato \&
Katsumi 2003) combined with  the
absolute magnitude (M$_{\rm V}$ = -5) for a F-type R CrB star in the LMC
(Alcock et al. 2001) puts V2552 Oph at a distance of about 7.7 kpc. At this distance,
interstellar gas following galactic rotation has a heliocentric
velocity of about -11 km s$^{-1}$. This velocity shifts to the
red for smaller distances: it is -2.3 km s$^{-1}$ for a distance of
1.1 kpc. The similarity between this range of predicted velocities
and the measured velocities
suggests that the Na D and K\,{\sc i} sharp components and the
diffuse interstellar bands at velocities of
-10 to 0 km s$^{-1}$  are interstellar
in origin. 

In addition to the interstellar components, there are sharp
components in the Na D lines at -38.4, -52.9, and a weak one
at -69.0 km s$^{-1}$.
These are possibly of circumstellar origin. If so, 
they represent clouds expanding away from the star at -98, -113, and
-129 km s$^{-1}$. These velocities exceed the surface escape velocity 
for a R CrB star. If it is assumed that the two stronger
components represent gas ejected at the time of the deep double
minimum reported by Kato \& Katsumi (2003), the radial
distances traveled are 93, and 135 stellar radii, respectively,
where a stellar radius is estimated at 80 solar radii. Such high
velocity ejections are commonly seen following a deep minimum.

\section{Circumstellar dust}

Declines of the R CrB stars are caused by soot clouds in front of
the star. A characteristic of the stars is an infrared excess
from circumstellar dust (Feast et al. 1997). Many R CrB stars
show an infrared excess at the K band. Most have
an infrared excess at longer wavelengths. A few have a strong excess at
60$\mu$m and 100$\mu$m from very cold dust in an extended shell
(Rao \& Nandy 1986; Walker 1986). 

To assess the infrared excess of V2552 Oph, we searched the 2MASS
All-sky Catalog of Point Sources 
(Cutri et al. 2003) for V2552 Oph's presence at J, H, and K. A source was
found coincident to within 0.4 arc sec with V2552 Oph's
position on an ESO R band frame. The 2MASS magnitudes
are J = 8.69, H = 8.39, and K = 8.16. The J-H and H-K color
indices, after correction for interstellar reddening corresponding to
E$_{\rm B-V}$ = 0.41, lie on the blackbody color-color plot at about
the temperature of the photosphere, i.e., there is no infrared
excess detectable at J, H, and K. V2552 Oph could not be located
in the {\it IRAS} Faint Source Catalog.

 V2552 Oph is one of few R CrB
stars not to show an excess in these bands.
In this respect, it is similar to Y Mus 
(Feast 1997). Kilkenny \& Whittet (1984) did find an
infrared excess for Y Mus at 5$\mu$m and 10$\mu$m. 
The absence for Y Mus of an infrared excess at K and
shorter wavelengths has been linked to the fact that the last 
minimum occurred several decades ago, i.e., recent soot production
has ceased or been reduced to low levels. This explanation is
not easily transferred to V2552 Oph which has experienced one or
two fadings recently. It is presumably nothing more than a
coincidence that V2552 Oph resembles Y Mus both in a lack of
infrared excess at K and in composition.

\section{Concluding Remarks}

Addition of V2552 Oph to the rare class of R CrB variables
is attributable to the Japanese amateur astronomer K. Haseda whose
announcement identifiying the star  as a variable included the
suggestion that it might be a R CrB variable.\footnote{
http://www.kusastro.kyoto-u.ac.jp/vsnet/Mail/alert6000/msg00226.html}
Our high-resolution spectrum fully confirms the identification of V2552 Oph
from photometry and low-resolution spectroscopy that the
star is a R CrB variable. We show that it is a member of the
majority class of these variables, as introduced by Lambert \&
Rao (1994). Within that class, V2552 Oph closely resembles
Y Mus and FH Sct in showing a below average oxygen abundance
and an above average abundance of Sr, Y, and Zr.
In this respect, V2552 Oph confirms the spectrum of compositions
offered by the majority stars but provides no novel clues to the
origins of the anomalous compositions of these remarkable
variables. Perhaps, the next discovered R CrB variable will
be the Rosetta Stone needed to unlock the chain tying the
abundances to the origins of these stars.

We thank B.E. Reddy for help with the acquisition and reduction of the
spectrum of V2552 Oph, G. Pandey for the model atmosphere,
 and D. Yong for the preparation of the
figures.

\clearpage
\clearpage

\plotone{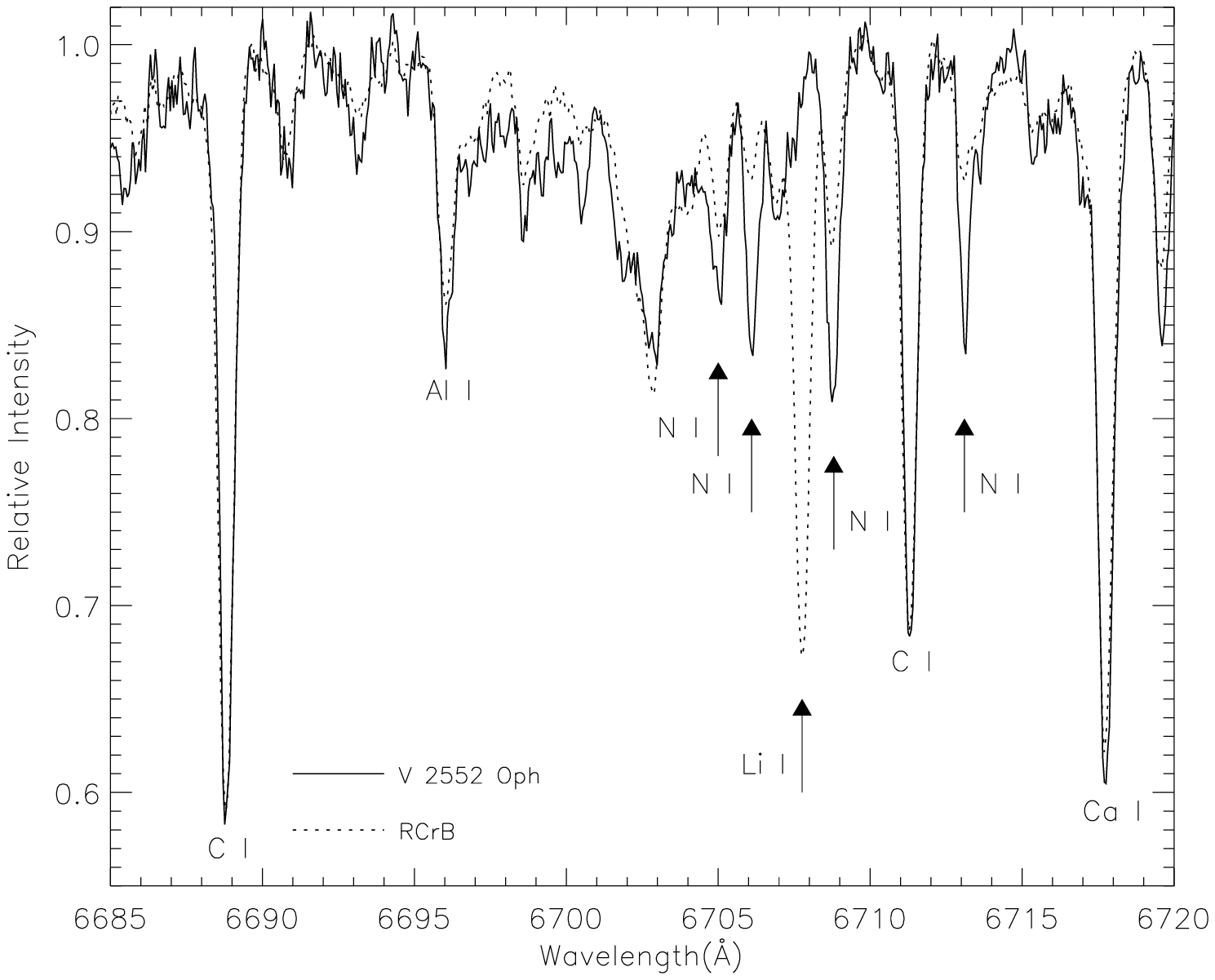}
\figcaption{Spectra of V2552 Oph and R CrB in the interval 6685-6720\AA.
Many lines are the same strength in the two spectra but remarkable
differences are seen and marked by an arrow: N\,{\sc i} lines are
much stronger in V2552 Oph but Li\,{\sc i} is strong in R CrB but absent from
V2552 Oph.}
\clearpage

\plotone{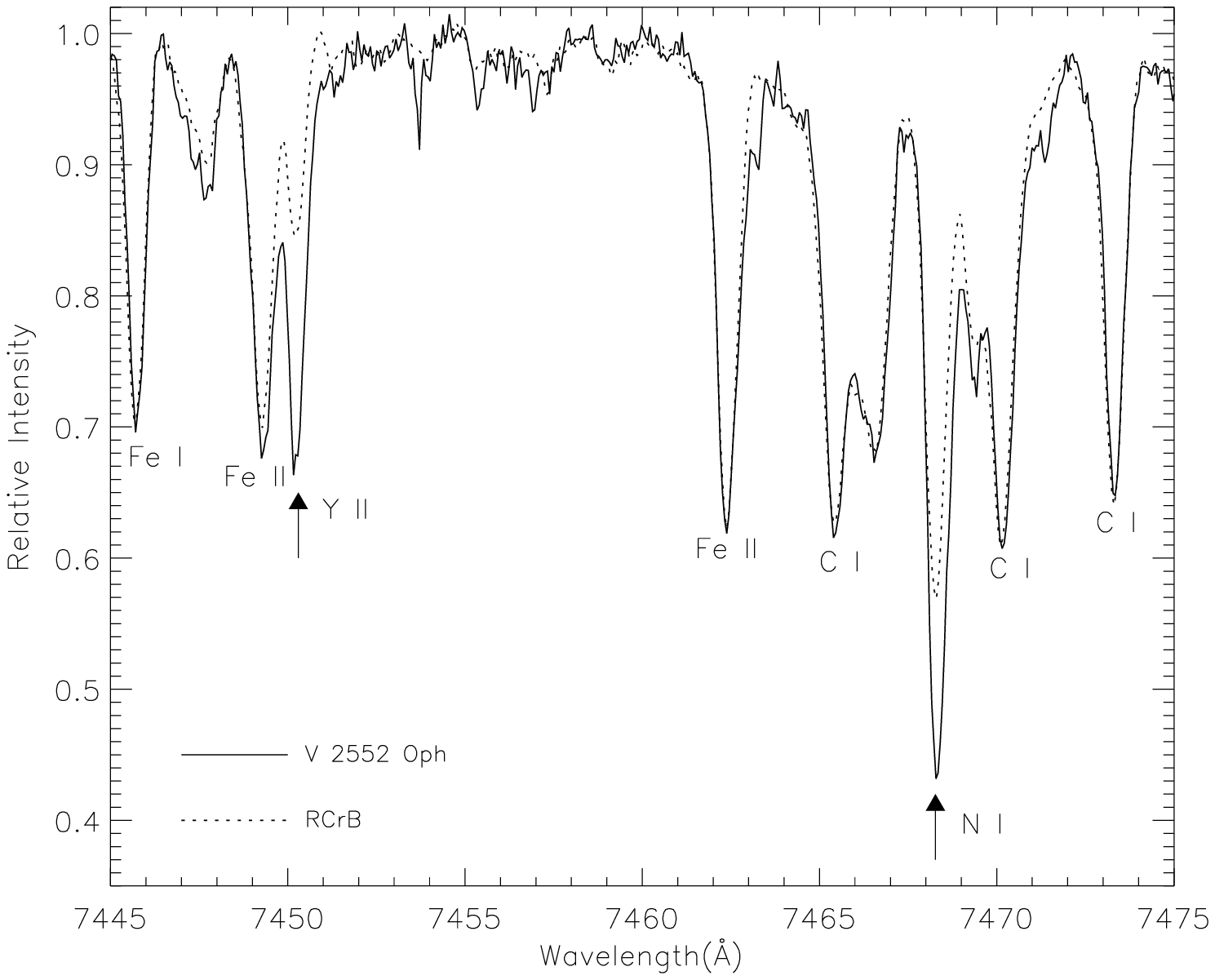}
\figcaption{Spectra of V2552 Oph and R CrB in the interval 7445-7475\AA.
Many lines are the same strength in the two spectra but remarkable
differences are seen and marked by an arrow: the N\,{\sc i} line
at 7468\AA\ and the Y\,{\sc ii} line at 7450\AA\ are
much stronger in V2552 Oph.}
\clearpage

\plotone{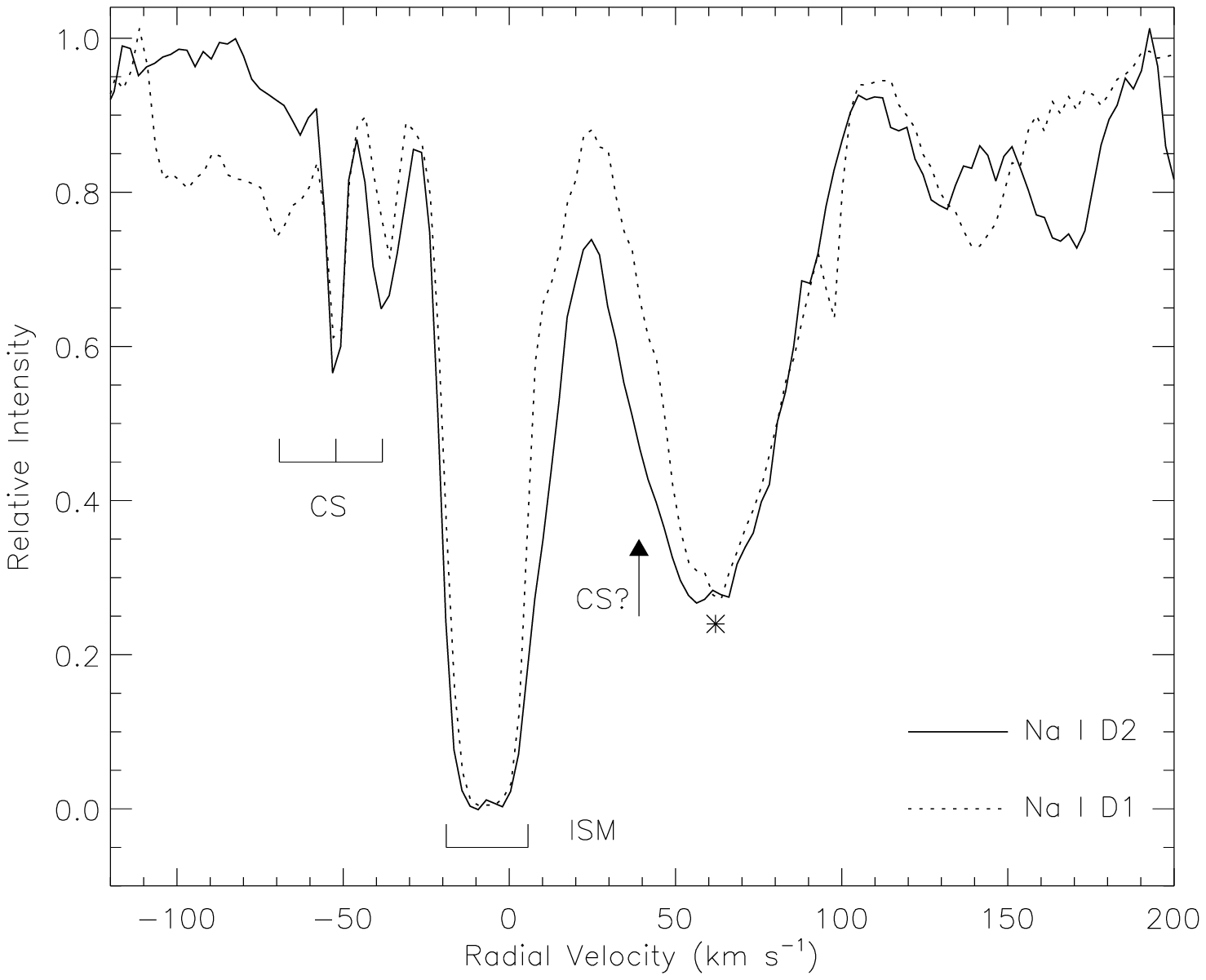}
\figcaption{The NaD1 and D2 lines in V2552 Oph. The stellar line at +60 km s$^{-1}$
is identified. A broad multi-component interstellar feature is indicated (see text).
High-velocity sharp circumstellar (CS) lines are marked. There is also a probable
CS component just to the blue of the stellar line and marked as CS?} 
\clearpage


\begin{center} \small TABLE I \\
  Elemental Abundances for V2552 Oph and R CrB
\end{center}

\small\begin{tabular}{lcllcr}
\hline\hline
  & R CrB &  & \underline{V2552 Oph} & \underline{n $^{a}$} &
\underline{Sun $^{b}$}
\\
\cline{2-3} Species  & Asplund et al. & present & & \\
\hline
H\,{\sc i} & 6.9 & 6.86 & 6.66 & 1, 1 & 12.00 \\
Li\,{\sc i} & 2.8 & 2.55  & \hspace{-.4cm} $<$0.96 & 1, 1 & 3.35 \\
C\,{\sc i} & 9.2 & 9.11 $\pm$ 0.25 & 9.05 $\pm$ 0.24 & 16, 17 & 8.46 \\
N\,{\sc i} & 8.4 & 8.42 $\pm$ 0.44 & 8.96 $\pm$ 0.28 & 5, 8 & 7.90 \\
O\,{\sc i} & 9.0 & 8.60 $\pm$ 0.37 & 7.96 $\pm$ 0.21 & 6, 6 & 8.76 \\
Na\,{\sc i} & 6.1 & 6.11 $\pm$ 0.05 & 6.14 $\pm$ 0.12 & 3, 3 & 6.37 \\
Mg\,{\sc i} &     & 6.81 $\pm$ 0.26 & 6.77 $\pm$ 0.09 & 2, 2 & 7.62 \\
Al\,{\sc i} & 5.8 & 5.76 $\pm$ 0.13 & 5.78 $\pm$ 0.11 & 2, 2 & 6.54 \\
Si\,{\sc i} & 7.2 & 6.97 $\pm$ 0.21 & 6.97 $\pm$ 0.22 & 4, 5 & 7.61 \\
Si\,{\sc ii} &    & 7.34 $\pm$ 0.15 & 7.67 $\pm$ 0.28 & 2, 2 &      \\
S\,{\sc i} & 6.8 & 6.70 $\pm$ 0.13 & 6.77 $\pm$ 0.09 & 7, 7 & 7.26 \\
K\,{\sc i} &     & 4.77 & 4.48 & 1, 1 & 5.18 \\
Ca\,{\sc i} & 5.3 & 5.32 $\pm$ 0.33 & 5.34 $\pm$ 0.27 & 5, 4 & 6.41 \\
Sc\,{\sc ii} &     & 2.89 $\pm$ 0.30 & 2.91 $\pm$ 0.23 & 3, 3 & 3.15 \\
Ti\,{\sc ii} &     & 4.05 $\pm$ 0.30 & 4.11 $\pm$ 0.37 & 2, 2 & 5.07 \\
Fe\,{\sc i} & 6.5 & 6.40 $\pm$ 0.24 & 6.37 $\pm$ 0.27 & 20, 21 & 7.54 \\
Fe\,{\sc ii} &    & 6.40 $\pm$ 0.13 & 6.42 $\pm$ 0.09 & 10, 10 &      \\
Ni\,{\sc i} & 5.5 & 5.49 $\pm$ 0.11 & 5.55 $\pm$ 0.18 & 4, 5 & 6.29 \\
Cu\,{\sc i} &     &                 & 3.95 $\pm$ 0.16 & 2, 2 & 4.34 \\
Zn\,{\sc i} &     & 4.16            & 4.31            & 1, 1 & 4.70 \\
Sr\,{\sc ii} &    &                 & 4.28            & 1    & 2.99 \\
Y\,{\sc ii} & 1.5 & 1.55 $\pm$ 0.18 & 2.29 $\pm$ 0.09 & 3, 3 & 2.28 \\
Zr\,{\sc ii} &    & 1.84 $\pm$ 0.19 & 2.25 $\pm$ 0.19 & 3, 3 & 2.67 \\
Ba\,{\sc ii} & 1.6 & 1.13 $\pm$ 0.16 & 1.03 $\pm$ 0.09 & 3, 3 & 2.25 \\
La\,{\sc ii} &     & 0.64 $\pm$ 0.38 & 0.63 $\pm$ 0.43 & 2, 2 & 1.25 \\
\hline\hline
\end{tabular}

$^{a}$\ n = number of lines used in the analyses: (a, b) where a and b are
number

\ \ of lines used for R CrB and V2552 Oph, respectively.

$^{b}$\ Recommended abundances for the solar system from Lodders

\ \ \  (2003, her Table 2).


\end{document}